\documentclass[11pt,aps,superscriptaddress,floatfix]{revtex4-2}

\usepackage[T1]{fontenc}
\usepackage{helvet}
\usepackage{charter}
\usepackage{graphicx}
\usepackage[dvipsnames]{xcolor}
\usepackage{url} 
\usepackage{hyperref}
\hypersetup{colorlinks,linkcolor=red,urlcolor=red,citecolor=blue}
\usepackage{dcolumn}
\usepackage{bm}
\usepackage{multirow}
\usepackage[symbol]{footmisc}
\usepackage{amsmath}
\usepackage{amssymb}
\usepackage{amsfonts}
\usepackage{subfigure}
\usepackage{url}
\usepackage{soul}
\usepackage[version=4]{mhchem}
\usepackage[normalem]{ulem}

\newcommand{\FG}{Fe$_3$Ga$_4$\,}

\begin{document}
	
	\title{Fluctuation-driven topological Hall effect in room-temperature \\ itinerant helimagnet \FG}
	
	\date{\today}
	
	\author{Priya R. Baral}
	\email{baralp@g.ecc.u-tokyo.ac.jp}
	\affiliation{Laboratory for Neutron Scattering and Imaging (LNS), PSI Center for Neutron and Muon Sciences, 5232 Villigen PSI, Switzerland}
	\affiliation{Institute of Physics, \'Ecole Polytechnique F\'ed\'erale de Lausanne (EPFL), CH-1015 Lausanne, Switzerland}
	\affiliation{Department of Applied Physics and Quantum-Phase Electronics Center, The University of Tokyo, Bunkyo-ku, Tokyo 113-8656, Japan}
	
	\author{Victor Ukleev}
	\affiliation{Laboratory for Neutron Scattering and Imaging (LNS), PSI Center for Neutron and Muon Sciences, 5232 Villigen PSI, Switzerland}
	\affiliation{Helmholtz-Zentrum Berlin f\"ur Materialien und Energie, D-14109 Berlin, Germany}
	
	\author{Ivica \v{Z}ivkovi\'{c}}
	\affiliation{Institute of Physics, \'Ecole Polytechnique F\'ed\'erale de Lausanne (EPFL), CH-1015 Lausanne, Switzerland}
	
	\author{Youngro Lee}
	\affiliation{Institute of Physics, \'Ecole Polytechnique F\'ed\'erale de Lausanne (EPFL), CH-1015 Lausanne, Switzerland}
	
	\author{Fabio Orlandi}
	\affiliation{ISIS Facility, STFC Rutherford Appleton Laboratory, Harwell Science and Innovation Campus, Oxfordshire OX11 0QX, United Kingdom}
	
	\author{Pascal Manuel}
	\affiliation{ISIS Facility, STFC Rutherford Appleton Laboratory, Harwell Science and Innovation Campus, Oxfordshire OX11 0QX, United Kingdom}
	
	\author{Yurii Skourski}
	\affiliation{Dresden High Magnetic Field Laboratory (HLD-EMFL), Helmholtz-Zentrum Dresden-Rossendorf, 01328 Dresden, Germany}
	
	\author{Lukas Keller}
	\affiliation{Laboratory for Neutron Scattering and Imaging (LNS), PSI Center for Neutron and Muon Sciences, 5232 Villigen PSI, Switzerland}
	
	\author{Anne Stunault}
	\affiliation{Institut Laue–Langevin, 71 avenue des Martyrs, CS 20156, Grenoble, 38042 Cedex 9, France}
	
	\author{J. Alberto Rodríguez-Velamazán}
	\affiliation{Institut Laue–Langevin, 71 avenue des Martyrs, CS 20156, Grenoble, 38042 Cedex 9, France}
	
	\author{Robert Cubitt}
	\affiliation{Institut Laue–Langevin, 71 avenue des Martyrs, CS 20156, Grenoble, 38042 Cedex 9, France}
	
	\author{Arnaud Magrez}
	\affiliation{Institute of Physics, \'Ecole Polytechnique F\'ed\'erale de Lausanne (EPFL), CH-1015 Lausanne, Switzerland}
	
	\author{Jonathan S. White}
	\affiliation{Laboratory for Neutron Scattering and Imaging (LNS), PSI Center for Neutron and Muon Sciences, 5232 Villigen PSI, Switzerland}
	
	\author{Igor I. Mazin}
	\affiliation{Department of Physics and Astronomy, George Mason University, Fairfax, VA 22030}
	\affiliation{Center for Quantum Science and Engineering, George Mason University, Fairfax, VA 22030}
	
	\author{Oksana Zaharko}
	\email{oksana.zaharko@psi.ch}
	\affiliation{Laboratory for Neutron Scattering and Imaging (LNS), PSI Center for Neutron and Muon Sciences, 5232 Villigen PSI, Switzerland}
	
	
	\maketitle
	\newpage
	
	\begin{center}
		\Large{ABSTRACT}
	\end{center}
	
	{\bf The topological Hall effect (THE) is a hallmark of a non-trivial geometric spin arrangement in a magnetic metal, originating from a finite scalar spin chirality (SSC). The associated Berry phase is often a consequence of non-coplanar magnetic structures identified by multiple \textbf{k}-vectors. For single-\textbf{k} magnetic structures however with zero SSC, the emergence of a finite topological Hall signal presents a conceptual challenge. Here, we report that a fluctuation-driven mechanism involving chiral magnons is responsible for the observed THE in a low-symmetry compound, monoclinic Fe$_3$Ga$_4$. Through neutron scattering experiments, we discovered several nontrivial magnetic phases in this system. In our focus is the helical spiral phase at room temperature, which transforms into a transverse conical state in applied magnetic field, supporting a significant THE signal up to and above room temperature.  Our work offers a fresh perspective in the search for novel materials with intertwined topological magnetic and transport properties.}
	
	\newpage
	
	\noindent\textbf{INTRODUCTION}
	
	The study of quantum materials with non-trivial topological properties has gained increasing popularity in condensed matter physics due to their potential applications in ultra-low-power electronic devices~\cite{tokura2019magnetic,nagaosa2020transport,tokura2017emergent}. The topological nature of these materials is determined by the emergent magnetic field arising from the unique geometric properties of the electronic band (spin) structure in reciprocal (real) space, which have the potential to support a locally enhanced Berry curvature. The anomalous Hall conductivity is a direct experimental manifestation of the topological contributions from these bands, appearing as an additional component to the normal Hall signal~\cite{nagaosa2010anomalous,verma2022unified}. Moreover, the overall Hall signal may exhibit a third component, known as the topological (or geometrical) Hall effect (THE), which arises from noncoplanar spin arrangements in real space. The latter is quantified by the so-called static scalar spin chirality (SSC), defined as $\chi_{ijk}=\mathbf{S}_i\cdot(\mathbf{S}_j\times\mathbf{S}_k)$ in the discrete (localized-spins) limit  with $\mathbf{S}_{i,j,k}$ being neighboring spins, forming a triangle electrons can hop around~\cite{nagaosa2013topological}. Equivalently, in the continuous limit of magnetisation $\mathbf m$ relatively slowly varying in space, the same quantity can be defined in terms of a vectors variable, called ``emergent magnetic field'', as  $\mathcal{B}_\alpha =1/2~\sum_{\beta\gamma} e_{\alpha\beta\gamma} \mathbf{m}\cdot(\partial_{\beta}\mathbf{m}\times\partial_{\gamma}\mathbf{m})$, where $\alpha,\beta$ and $\gamma$ are Cartesian coordinates in real space. This emergent field deflects conduction electrons and produces a finite THE, among other emergent phenomena~\cite{jiang2017direct,kang2016skyrmion,tokura2017emergent,hirschberger2020topological}.
	
\begin{figure*}[htb!]
		\centering
		\includegraphics[width=1.0\linewidth]{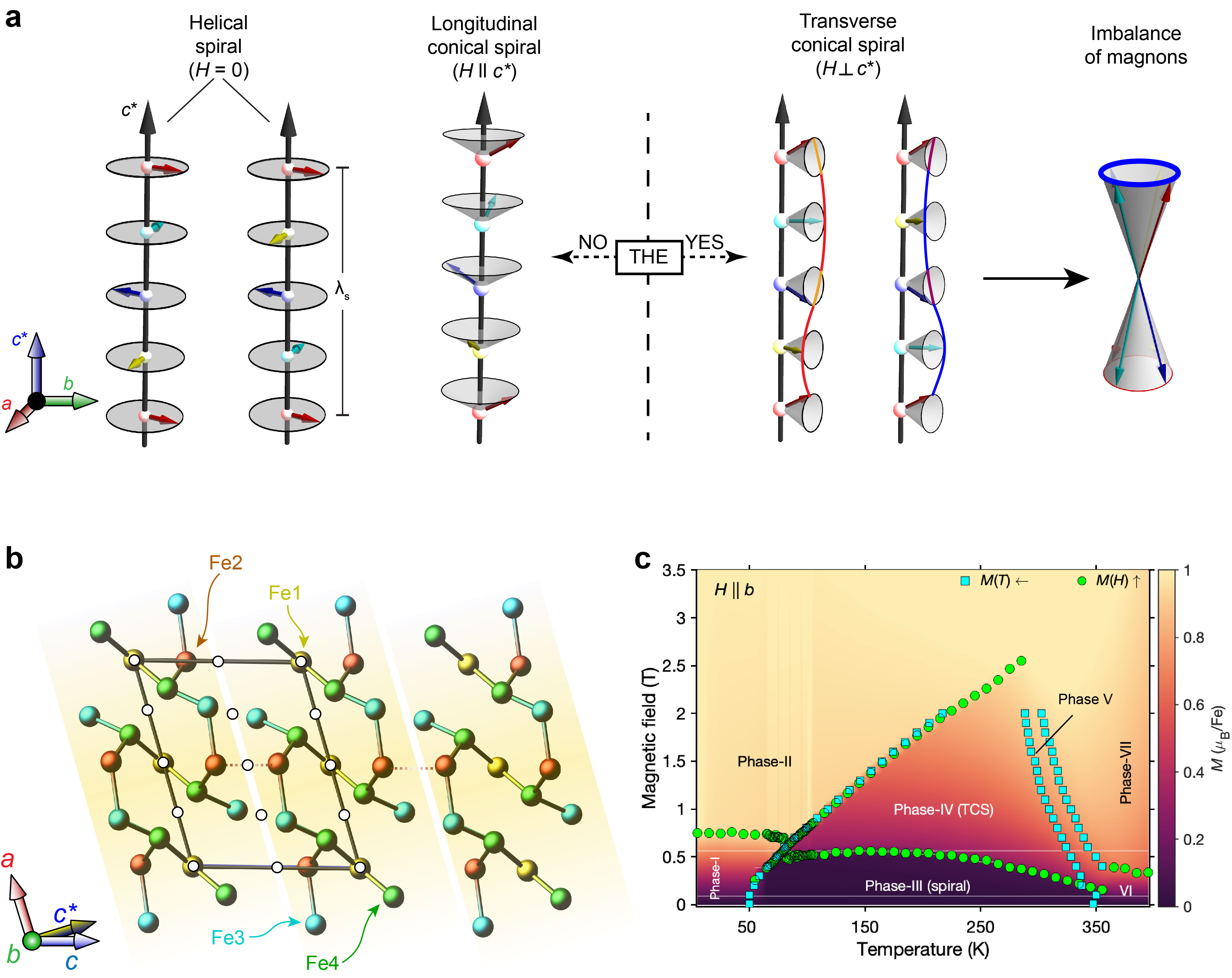}
		\caption{\textbf{Magnetic field induced dynamical scalar spin chirality in \FG.}{\bf a,} Schematics of the two counter-rotating spirals propagating along $c^{\ast}$ and $-c^{\ast}$, respectively. A longitudinal (transverse) conical spiral state is formed when magnetic field is applied along (orthogonal to) the direction of spiral propagation. Preferential excitation of magnons with a particular handedness generates a dynamic scalar spin chirality. The topological Hall effect could exist in the cases shown to the right and could not in the cases to the left. {\bf b}, Unit cell of \FG~projected onto the $ac$-plane showing four inequivalent Fe$_{1-4}$-sites (filled colored circles) and the inversion centers (empty black circles). The overall structure can be conceived as Fe-slabs stacked along the $c$-axis, schematically shown by yellow shaded regions. Two different types of the Fe-Fe paths (shorter than 2.95 \AA) are drawn: the intra-slab couplings are shown by solid lines, while the dashed lines represent the inter-slab couplings along the $c$-axis. {\bf c,} Magnetic phase diagram constructed from temperature and field dependent magnetisation measurements. The color map corresponds to the magnetisation evolution in various phases.}
		\label{Fig1}
	\end{figure*}

	Not all noncoplanar structures generate a finite scalar spin chirality. For instance, a conical spiral (see Fig.~\ref{Fig1}a), described by the equation $d\mathbf{m}/dz=\mathbf{\Omega}\times \mathbf{m}+\mathbf{m'}$, where $z$ is the direction of the spiral propagation, and $\mathbf{\Omega}$ and $\mathbf{m'}$ are arbitrary vectors, obviously has zero emergent field, even in the noncoplanar case of $\mathbf{m'}\cdot\mathbf{\Omega}\neq 0$, because for a nonzero $\mathcal{B}$ one needs to have magnetisation varying along two independent directions. Similarly, a combination of two helical spirals can also not generate $\mathcal{B}\neq 0$. Indeed, one can show that in this case $\mathcal{B} =1/2~m^2\mathbf{m}\cdot(\mathbf{\Omega_1}\times\mathbf{\Omega_2})$, which averages to zero.
	
	However, it was realized in recent years that multiple spirals can generate THE, if they are properly combined. This includes three independent spirals (and each of them can be flat) ~\cite{neubauer2009topological,kanazawa2015discretized,kurumaji2019skyrmion,hirschberger2019skyrmion}, or two spirals combined with a net uniform magnetisation (as indicated by the equation above) ~\cite{khanh2020nanometric,takagi2022square,yoshimochi2024multi}, as well as other noncoplanar arrangements that cannot be described as combinations of spirals ~\cite{kanazawa2011large,fujishiro2019topological}. 
	
	Interestingly, significant topological Hall signals have been recently observed in a topologically trivial magnetic phase of kagome-layered YMn$_{6}$Sn$_{6}$, known as the transverse conical spiral (TCS) state, which formally carries zero scalar chirality and emergent field~\cite{ghimire2020competing}. As schematically shown in Fig.~\ref{Fig1}a, the TCS state is obtained after a spin-flop while applying an in-plane magnetic field to the helical spiral and can be visualized as a cycloid propagating orthogonally to the applied magnetic field, together with a uniform magnetisation along the field. The appearance of a THE in YMn$_{6}$Sn$_{6}$ is attributed to a {\em dynamical} effect, where directional  breaking of time-reversal symmetry leads to an unbalanced population of chiral magnons with a given handedness, i.e. a nematic spin chirality. The resulting susceptibility generates the necessary SSC to promote the THE signal at elevated temperatures, and the THE amplitude is roughly proportional to the temperature. Experimental realisations of fluctuation-driven THE are currently scarce and largely restricted to the family of hexagonal $R$Mn$_{6}$Sn$_{6}$ compounds~\cite{ghimire2020competing,zhang2022magnetic,fruhling2024topological}, though the theory is generic and is not limited to any symmetry of the lattice.
	
	In this connection, it would be intriguing to discover other non-chiral materials that nevertheless exhibit topological Hall effect arising due to the fluctuation-driven mechanism. To this end, we turn our attention to the itinerant magnet \FG. We construct a more detailed magnetic phase diagram with additional phases compared to the ones currently found in literature by combining high-resolution magnetometry and electrical transport data~\cite{mendez2015competing,samatham2017weak}. Our neutron diffraction results, combined with spherical neutron polarimetry (SNP), provide strong support for the helical spin arrangement at elevated temperatures with moments rotating in the $ab$-plane, which flops into a TCS state in magnetic fields applied along the $a$- or $b$-axis. We propose that the observed THE results from an unbalanced excitation of chiral magnons in an applied magnetic field, suggesting a dynamic origin of the SSC. This mechanism is conceptually similar to that in YMn$_{6}$Sn$_{6}$, but occurs in a completely different magnetic material. The topological Hall signal, which is proportional to the emergent magnetic field generated by this SSC, varies linearly with temperature at a fixed magnetic field. This observation strengthens our claims about its dynamic origin. The results obtained for the TCS phase as well as for the zero field helical spiral phase are discussed in conjunction with other regimes of the thoroughly explored phase diagram. The magnitude of the topological Hall signal is comparable to that measured in topologically non-trivial skyrmion hosts. Our findings demonstrate an alternative route for generating a geometrical Hall response in itinerant spiral magnets at elevated temperatures.\\
	
	\noindent\textbf{RESULTS}
	
	\noindent\textbf{Crystal structure and complex magnetic phase diagram of \FG}
	
	\noindent Our target material crystallises in a centrosymmetric structure shown in Fig.~\ref{Fig1}b with the monoclinic unit cell ($C 2/m$, $a$ = 10.0966(5) \AA, $b$ = 7.6650(4) \AA, $c$ = 7.8655(4) \AA, and $\beta$ = 106.25(4)$^{\circ}$ at $T$~=~300~K~\cite{philippe1975structures}). Results from detailed structural characterisation using single crystal X-ray diffraction can be found in supplementary materials section~{\color{red}I}. The complicated network of four inequivalent Fe-sites can be separated into Fe-slabs stacked along the $c$-axis. Within each slab the nearest iron distance does not exceed 2.95~\AA~and majority of the Fe-bonds contain no inversion center at the middle, allowing for local Dzyaloshinskii–Moriya interaction (DMI). The Fe slabs have not only geometrical, but also electronic relevance for magnetic properties discussed below, since the DFT spin spiral calculations~\cite{afshar2021spin} point to the energy minimum for magnetic spirals with propagation vectors along the \textbf{Q} = (0, 0, $q_z$) reciprocal lattice vector.
	
\begin{figure*}[htb!]
		\centering
		\includegraphics[width=1.0\linewidth]{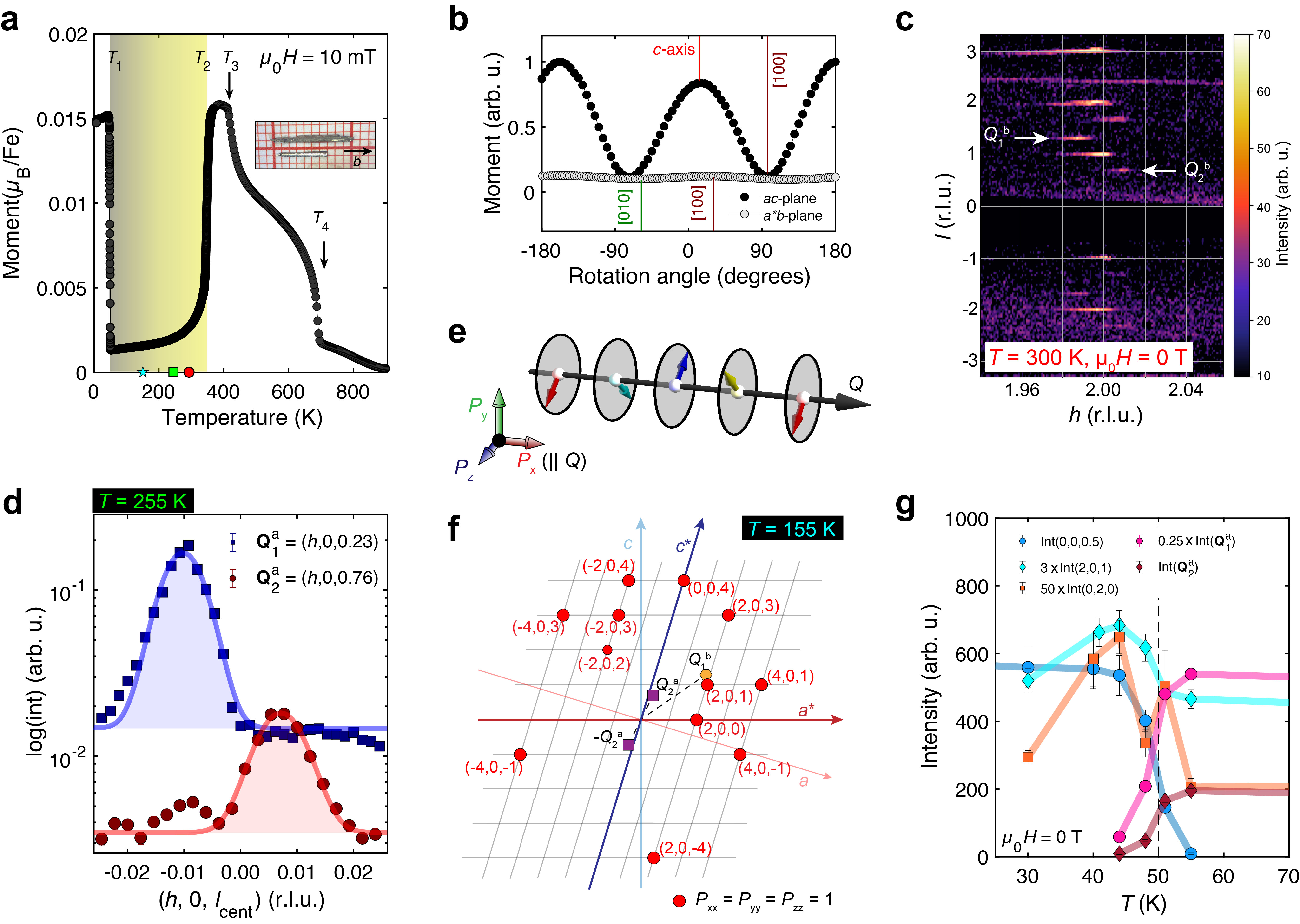}
		\caption{\textbf{Non-coplanar helimagnetism in \FG\,at room temperature.} {\bf a,}  Temperature dependence of magnetisation between 5~K and 900~K in 10~mT magnetic field applied along the $b$-axis with four transitions labelled $T_1 - T_4$. Helimagnetism observed in Phase III is schematically shown as a shaded region between $T_1$ and $T_2$. Inset shows typical \FG single crystals used in this study. Cyan, green and red symbols represent the corresponding temperatures for spherical neutron polarimetry (SNP) on D3, single crystal diffraction on DMC and WISH instruments, respectively. {\bf b,} The measured $M$($\psi$) curves for a rotating single crystal under a constant magnetic field of 100~mT applied normal to the plane of measurement at 155~K. {\bf c,} Orthogonal projection of the monoclinic ($h$ 0 $l$) reciprocal lattice plane at $T$~=~300~K obtained from data collected on the WISH diffractometer. The small deviations from the ($h$,0,$l$) line ($h$ = 2) for the magnetic satellites indexed with $\mathbf{Q}_1^{\mathrm{b}}$ and $\mathbf{Q}_2^{\mathrm{b}}$ are apparent. {\bf d,} Order of magnitude difference in the intensity of  $\mathbf{Q}_1^{\mathrm{a}}$ and  $\mathbf{Q}_2^{\mathrm{a}}$ obtained at 255~K. {\bf e,} Local coordinate axes in the SNP experiment, where $P_{x}$ component of the incoming neutron beam is aligned along the $\mathbf{Q}$-direction of each reflection. {\bf f,} All reflections probed with the SNP experiment at 155~K. {\bf g,} Temperature dependence of intensity of several Bragg peaks tracked in the vicinity of $T_1$ while warming.}
		\label{Fig2}
	\end{figure*}

	The complex magnetic phase diagram in Fig.~\ref{Fig1}c contains numerous phases which we identify by magnetisation and ac susceptibility measurements. As shown later in this article, various scattering techniques together with transport measurements have been used to understand the macroscopic features of the identified phases. Magnetisation data presented in Fig.~\ref{Fig2}a indicates magnetic ordering of \FG~at temperature $T_4$ around 720~K. On cooling three successive magnetic transitions could be distinguished, at $T_3 \simeq$ 420~K, $T_2 \simeq$ 370~K and $T_1 \simeq$ 53~K signaling the onset of Phases VI, III and I, respectively. The existence of the previously unreported Phase V is evident from the additional magnetometry data presented in supplementary materials Fig.~{\color{red}S2b}. As discussed later in the article, we also observe strong signatures at the boundaries of Phase V in the electrical transport measurements.\\
	
	\noindent\textbf{Non-coplanar magnetic order at zero field in Phase III}
	
	A THE signal was previously detected in applied magnetic field within the temperature range 68 K <$T$< 360 K~\cite{mendez2015competing}, which corresponds to Phase IV in our magnetic phase diagram shown in Fig.~\ref{Fig1}c. Thus, our discussion starts with the nature of magnetic order arising in the zero field state within the same temperature range, that is Phase III. As shown in Fig.~\ref{Fig2}a, in presence of a small magnetic field, the occurrence of this phase is signaled by a strong gradual reduction in the magnetic moment at the high-$T$ side, $T_2$, and by an order of magnitude sharp increase in the net magnetisation at the low-$T$ side, $T_1$. To characterise magnetic anisotropy in Phase III we measure the change of magnetisation within the $ac$- and $a^{\ast}b$-planes (Figure~\ref{Fig2}b). Apparently, magnetisation follows the order $m_c \gg m_a \geq m_b$ suggesting almost equal moment distribution in the $ab$-plane. A possible interpretation consistent with this scenario could be a spiral propagating along $c^{\ast}$ direction, whose local moments are confined within the $ab$-plane.
	
	A previous neutron diffraction study showed that phase III is incommensurate (ICM), with the propagation vector \textbf{k}~=~(0~0~$\gamma$), where $\gamma$ is temperature dependent and varies between $\sim$0.24 and $\sim$0.29 r.l.u.\cite{wu2018spin}. Two contradicting models for the ICM magnetic order are proposed: an amplitude modulated spin density state (ASDW) and a helical state, both propagating along $c^{\ast}$. While in the ASDW model magnetic moments are confined within the $ac$-plane with only the Fe$_4$ moments possessing a small $b$-component~\cite{wu2018spin}, the magnetic moments rotate in the $ab$-plane in the helical model~\cite{afshar2021spin, wilfong2022helical}. 
	
	In order to clarify the nature of magnetic order in Phase III we employ a combination of single crystal neutron diffraction (SND) and SNP experiments (see Methods for further details). A snapshot of the reciprocal space map obtained with high-resolution on the time-of-flight diffractometer WISH at ISIS is shown in Fig.~\ref{Fig2}c. Besides the ICM $c$* component reported earlier by Wu \textit{et al.}~\cite{wu2018spin}, we also detect a small component along $h$ (Fig.~\ref{Fig2}c \& d). Thus the ICM reflections should be properly indexed as ($h+\alpha$, 0, $l+\gamma$). This does not lower the magnetic symmetry however, since ($h+\alpha$, 0, $l+\gamma$) and (0, 0, $l+\gamma$) belong to the same plane of symmetry in the first Brillouin zone. Both $\alpha$ and $\gamma$ change with temperature and field. At $T$~=~255~K for a low$-\mathbf{Q}$ reflection $\mathbf{Q}_2^{\mathrm{a}}$, $\alpha$ and $\gamma$ are determined to be 0.007 and -0.240, respectively, see Fig.~\ref{Fig2}d. The other low$-\mathbf{Q}$ satellite $\mathbf{Q}_1^{\mathrm{a}}$ inside the first Brillouin zone can be indexed with -$\alpha$ and -$\gamma$ within the precision of the measurement, which suggest presence of two chiral domains, this is elaborated further below.
	
	In the next step, the directional information about the magnetisation distribution in Phase III is deduced from our SNP experiment. The experimental setup is schematically shown in Fig.~\ref{Fig2}e and Fig.~{\color{red}S3}, where the direction of the incident neutron polarisation is fixed relative to $\mathbf{Q}$. The measured polarisation matrices $P_{ij}$ ($i$- the row-index of incoming polarisation, $j$- the column index for outgoing polarisation) are presented in supplementary materials. The analysis of three components of the scattered beam makes SNP a powerful tool to study non-coplanar magnetic orders.
	
	Several reflections were measured in the ($h0l$) scattering plane, see the full map in Fig.~\ref{Fig2}f. For the reflection $\mathbf{Q}_2^{\mathrm{a}}$=(0.007, 0, 0.759) we obtain finite chiral matrix elements: $P_{yx}$ = -0.43(5) and $P_{zx}$ = -0.48(5). This establishes a helical nature of the magnetic order. $P_{xx}$ equals -1.04(3), implying a purely magnetic origin, while the remaining elements are zero. The polarisation matrix for the $\mathbf{Q}_1^{\mathrm{b}}$=(1.99, 0, 1.28) reflection also has finite chiral elements, but of the positive sign, opposite to $\mathbf{Q}_2^{\mathrm{a}}$. From the finite and opposite sign $P_{yy}$, $P_{zz}$ elements for the $\mathbf{Q}_1^{\mathrm{b}}$ reflection we extract that the magnetic component along the local $z$- (crystal $b$-) axis is larger than the magnetic component along the $y$- (approximately (101)- crystal) axis. In summary, Phase III is unambiguously determined to be helical and chiral domains are unequally populated allowing for a net spin chirality. In addition, the magnetic arrangement  might be three-dimensional. Regrettably, a reliable collection of intensity data-set was not possible due to an insufficient quality of crystals at this low temperature, therefore the microscopic model of Phase III cannot be developed further. Still, in accordance with the magnetic anisotropy results, the moments should be located predominantly in the $ab$-plane. 
	
	As the temperature decreases, around $T_1$, we observe a narrow coexisting region between Phase III and Phase I (see Fig.\ref{Fig2}g). In Phase I, in addition to the ferromagnetic component ($\mathbf{k}$ = 0) reported in Refs.~\cite{wu2018spin, mendez2015competing}, we detect a $\mathbf{k}$ = (0~0~1/2) propagation vector and associated antiferromagnetic commensurate order, which was previously overlooked. Phase I and its transformation into Phase II in magnetic field are discussed in details in Supplementary material.\\
	
	\noindent\textbf{Magnetic field-induced effects in \FG}
	
	\begin{figure*}[htb!]
		\centering
		\includegraphics[width=1.0\linewidth]{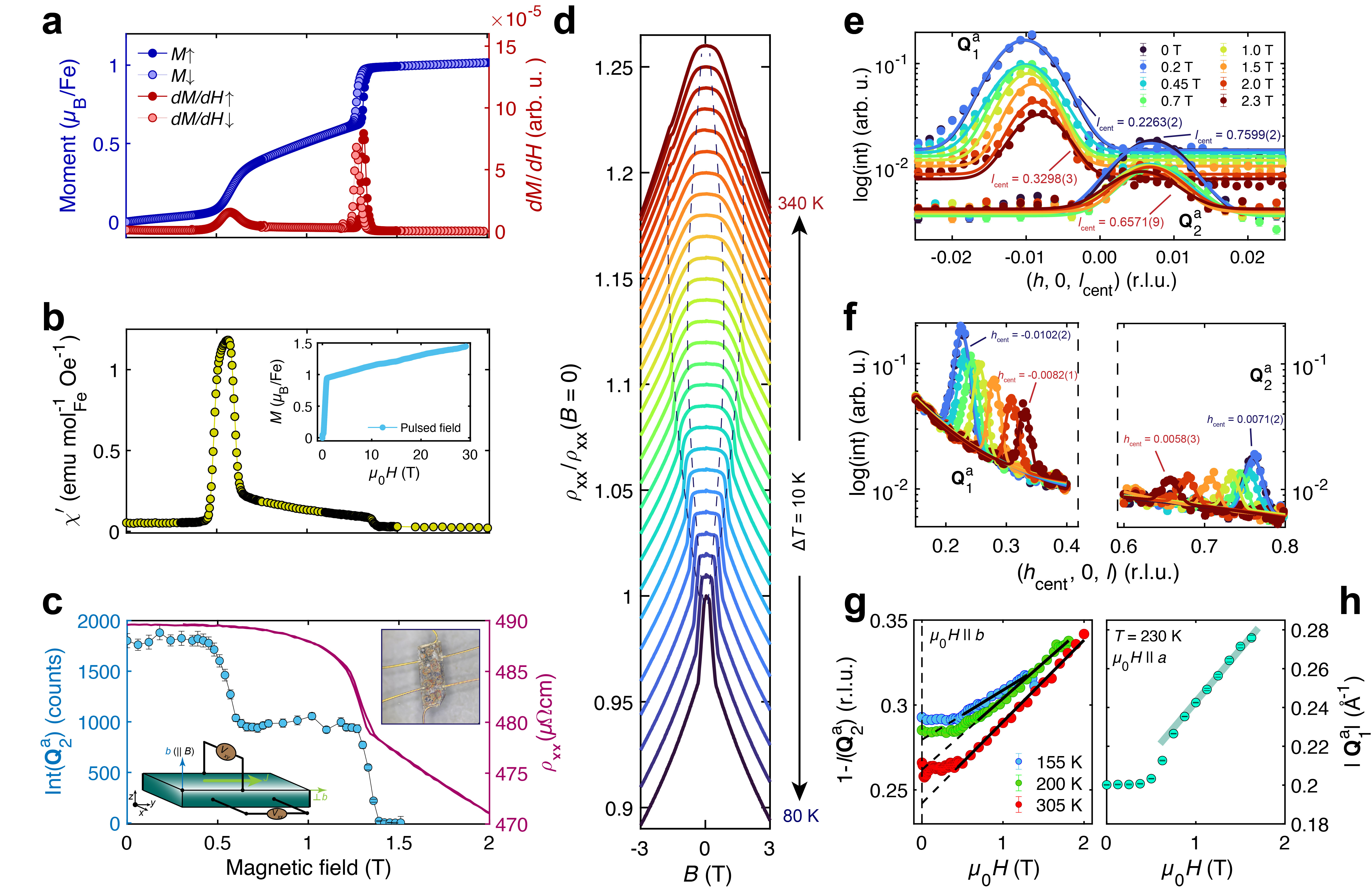}
		\caption{\textbf{Helical spiral transformation for in-plane magnetic field.} {\bf a,} Magnetic moment as a function of applied field at $T$ = 155~K is compared with the corresponding $dM$/$dH$ curve. A small hysteresis is observed at $H_{\mathrm{c2}}$. {\bf b,} Real part of the ac susceptibility ($\chi^{\prime}$) data shows a clear transition at $H_{\mathrm{c1}}$ and a weak step-like feature at $H_{\mathrm{c2}}$. The inset shows increasing magnetisation in the $M$($H$) curve obtained at the pulsed field facility up to 30~T at $T$ = 10 K. {\bf c,} Magnetic field dependence of the $\mathbf{Q}_2^{\mathrm{a}}$ = (0.01, 0.00, 0.72) intensity and the longitudinal resistance ($\rho_{xx}$) at $T$~=~155~K. Inset shows the plate-type crystal used for electrical transport measurements. {\bf d,} Longitudinal magnetoresistance measured in Phase III of \FG between 80~K and 340~K at every 10~K as a function of internal field $B$. The dashed lines represent phase boundaries corresponding to $H_{\mathrm{c1}}$ and $H_{\mathrm{c2}}$. A constant offset of 0.01 is applied between successive curves for better visualisation. Magnetic field variation of the {\bf e,} $h$- and {\bf f,} $l$- components of the $\mathbf{Q}_1^{\mathrm{a}}$ and $\mathbf{Q}_2^{\mathrm{a}}$ ICM reflections. {\bf g,} and {\bf h,} show the $q_z$ changes with respect to applied magnetic field for the $\mathbf{Q}_2^{\mathrm{a}}$ and $\mathbf{Q}_1^{\mathrm{a}}$ reflections, respectively. In the region between $H_{\mathrm{c1}}$ and $H_{\mathrm{c2}}$ the $q_z$ changes were fitted  for all three temperatures with straight lines, they were extended (dotted lines) down to $\mu_0H$ = 0~T. Data shown in Panels-(e) to (g) were measured at $T$ = 255~K, whereas for panel-h $T$ = 230~K.} 
		\label{Fig3}
	\end{figure*}
	
	Isothermal magnetisation and susceptibility measured at $T$ = 155 K (Figs.~\ref{Fig3}a, b) disclose two transitions induced by magnetic field applied along the $b$-axis: at $H_{\mathrm{c1}}$ = 0.55~T and $H_{\mathrm{c2}}$ = 1.4~T. The step-like increase in magnetisation are clear signatures of the successive transitions from Phase III to Phase IV and then to Phase II. Around $H_{\mathrm{c2}}$, there is a clear hysteresis revealing the first-order nature of the transition between these phases. As shown in Fig.~\ref{Fig3}a, at $T$~=~155~K there is no hysteresis around $\mu_0 H=0$~T. This is corroborated by our SNP results, since full polarisation matrices for the measured $\mathbf{k}$~=~0 reflections suggest only nuclear, no magnetic contributions (see Supplementary material). These results indicate the absence of any ferromagnetic component in Phase III of \FG.
	
	Next we follow neutron diffraction signatures of the magnetic field induced phases emerging out of Phase III. As shown in Fig.~\ref{Fig3}c, intensity of the $\mathbf{Q}_2^{\mathrm{a}}$ reflection is approximately halved in Phase IV compared to the zero field value. The reduction is abrupt at $H_{\mathrm{c1}}$ and $H_{\mathrm{c2}}$ and it correlates with the derivative of magnetisation $\Delta M$/$\Delta H$, while within Phases III and IV the intensity is constant. This behaviour suggests the moment reorientation from the $ab$- to $ac$- plane, hence the transformation of the incommensurate modulation from a helix to a cycloid. The magnetic field-induced phase changes also have subtle manifestation in the longitudinal resistivity data measured throughout Phase IV, as shown in Fig~\ref{Fig3}c \& d, indicating a discontinuous transformation between Phases IV and II. A weak (strong) change observed in $\rho_{xx}$ at $H_{\mathrm{c1}}$ ($H_{\mathrm{c2}}$) indicates a rather weak (strong) charge-spin coupling at the corresponding transitions.
	
	The positional shift of ICM reflections is also strong but continuous with field (Fig.~\ref{Fig3}e \& f). For the $\mathbf{Q}_1^{\mathrm{a}}$ reflection at $T$ = 255~K the $l$-value increases from 0.2263(5) at $\mu_0H$ = 0~T to 0.3298(3) at $\mu_0H$ = 2.3~T resulting in $\Delta l$/$\overline l$= 18.6\%. For the $\mathbf{Q}_2^{\mathrm{a}}$ reflection $\Delta (1-l)$/$\overline{(1-l)}$=17.6\%, and the $h$-values change by a similar amount (Fig.~\ref{Fig3}e). Fig.~\ref{Fig3}g and h present the field evolution of the $q_z$-component in Phase IV for the $\mathbf{Q}_2^{\mathrm{a}}$ and $\mathbf{Q}_1^{\mathrm{a}}$ reflections, respectively. The field dependence is linear and the $l$($H$) slope is positive revealing the tendency towards a commensurate structure with $q_z$=1/3, as schematically depicted in Fig.~{\color{red}S4}. Further field dependence of $\mathbf{Q}_1^{\mathrm{a}}$ reflection is depicted in supplementary material section~{\color{red}V}. However, as shown in the supplementary material, the $q_z$=1/2 magnetic order develops at low temperatures and elevated magnetic fields, suggesting a competition between the $q_z$=1/3 and $q_z$=1/2 propagation vectors.
	
	\begin{figure*}[htb!]
		\centering
		\includegraphics[width=1.0\linewidth]{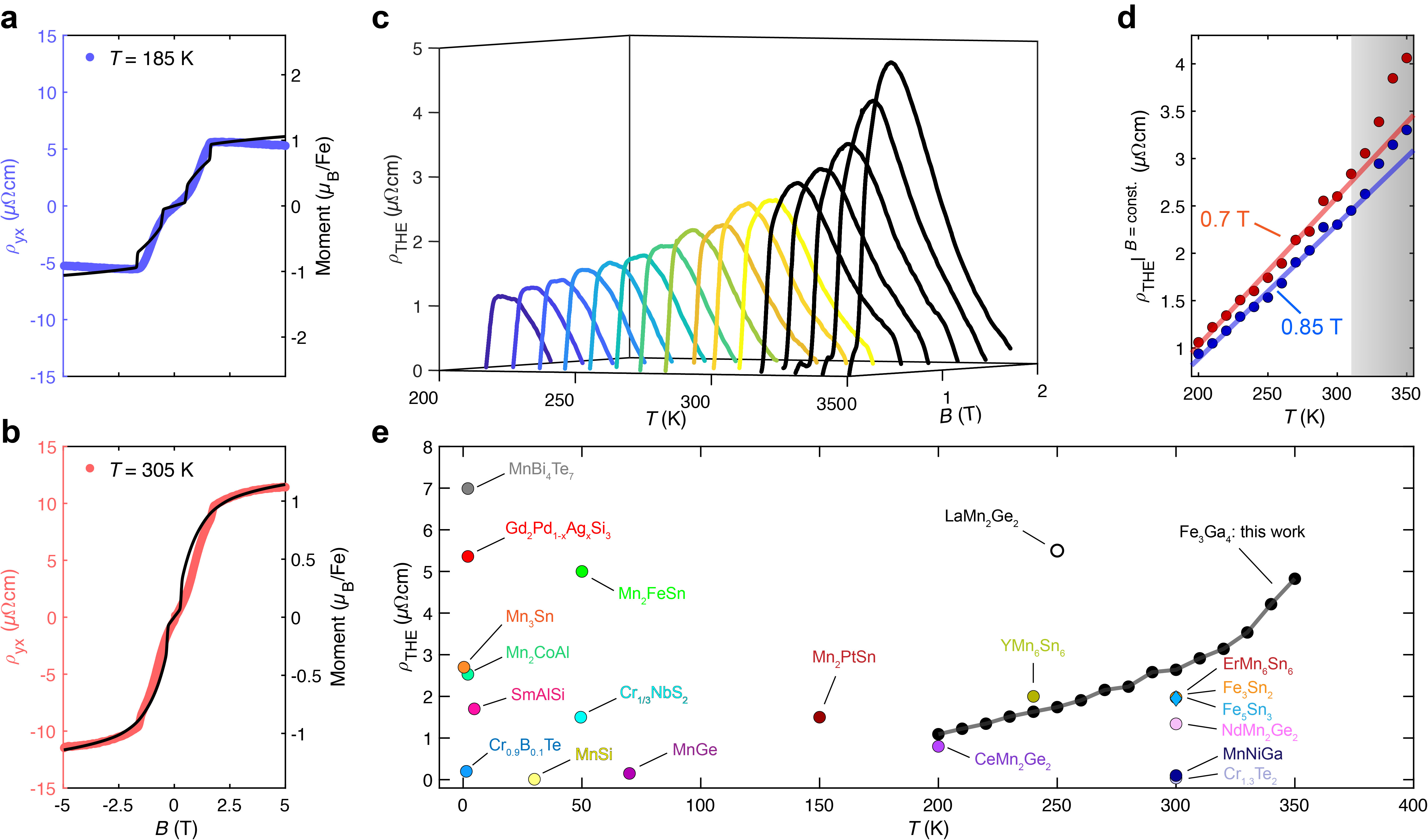}
		\caption{\textbf{Fluctuation-induced topological Hall effect in the transverse conical spiral state.} Comparison between Hall resistivity ($\rho_{yx}$) and magnetisation data at {\bf a,} 185 K and {\bf b,} 305 K. {\bf c,} Geometrical contribution ($\rho_{\mathrm{THE}}$) to the total Hall resistivity shown as a function of magnetic field between 200~K and 350~K, after subtracting the normal and anomalous contributions. {\bf d,} $\rho_{\mathrm{THE}}$ as a function of temperature at two constant magnetic fields. The data were fitted with a linear function in Phase IV. Deviation from the linear behavior, marked by shaded region signifies the appearance of Phase V. {\bf e,} Comparison between maximum of topological Hall resistivity ($\rho_{\mathrm{THE}}$) in various bulk magnets, among which some even host, as opposed to \ce{Fe3Ga4}, skyrmion lattices. Data shown in panel-e are taken from Refs~\cite{surgers2014large,hirschberger2020topological,liu2022large,wang2016centrosymmetric,roychowdhury2021giant,wang2020giant,zheng2021giant,zhang2023room,liu2018giant,yao2023large,rout2019field,xu2022giant,hall2022comparative,he2020large,li2020large,fruhling2024topological,kanazawa2011large,wang2021field,neubauer2009topological,roychowdhury2024giant}.}
		\label{Fig4}
	\end{figure*}
	
	Next we evidence the existence of a finite topological Hall resistivity ($\rho_{\mathrm{THE}}$) through a comparison between the transverse component of resistivity data ($\rho_{yx}$) with corresponding isothermal magnetisation data presented in Fig.~\ref{Fig4}a \& b. The extraction of $\rho_{\mathrm{THE}}$ is explained in more details in the supplementary material. As shown in Fig.~\ref{Fig4}c, we observe a proportionate increase in $\rho_{\mathrm{THE}}$ at elevated temperatures, corresponding to growing fraction of Phase IV in our magnetic phase diagram (see Fig.~\ref{Fig1}c). 
	
	From the quantitative assessment shown in Fig.~\ref{Fig4}d, we observe a $T$-linear increase in |$\rho_{\mathrm{THE}}$|, strongly supporting the fluctuation-based mechanism of THE. The magnitude of $\rho_{\mathrm{THE}}$ observed in \FG matches that arising from the short-periodic topologically protected spin textures, such as a skyrmion lattice (SkL) phase~\cite{kurumaji2019skyrmion,hirschberger2020topological}. But, unlike in these magnets, THE observed in \FG is generated by a single-$\mathbf{k}$ incommensurate magnetic structure, at room temperature.\\
	
	\newpage
	
	\noindent\textbf{\large DISCUSSION}
	
	\noindent 
	Our experimental findings on \FG, particularly the SND and SNP results, reveal crucial details of both high-temperature Phases III, IV, as well as low-temperature Phases I, II. This allows us to speculate about the main components of the exchange Hamiltonian of the system and to propose the origin of the observed THE in Phase IV. 
	
	The propagation vector of Phase III has two incommensurate components, $q_z$ along $c^*$ ($\gamma\approx$ 0.27) and  $q_x$ along $a^*$  ($\alpha\approx$ -0.01). The experimental behaviour of the $\gamma\approx$ 0.27 component in Phase III is already highlighted in the previous section, here we discuss its theoretical validation. Non-relativistic spiral spin calculations~\cite{afshar2021spin} favor magnetic orders with propagation vectors along the (0 0 $q_z$) direction over the general ($q_x~q_y~q_z$) solutions, with a shallow energy minimum at $q_z$= 0.27. These calculations predict that the observed commensurate propagation vector $q_z$= 0.5 of Phase I is higher in energy than the incommensurate vector of Phase III with $q_z$= 0.27. However, antiferromagnetic cell doubling is much more favorable than the ferromagnetic ground state proposed in Ref.~\cite{wu2018spin}. Spiral calculations are not possible with spin-orbit coupling (SOC), since the generalized Bloch theorem is not applicable~\cite{sandr1986}. Therefore, we performed calculations with SOC in a quadrupled supercell, using the same computational setup as in Ref.~\cite{afshar2021spin}, thus accessing three energies: for $q_z=$ 0, 0.25 and 0.5. The results are shown in Figure~{\color{red}S5}.	The energy difference between the two latter cases is reduced from 1.12 meV/Fe to 0.45 meV/Fe ($\approx 5$ K). Still, the calculated energy of the $q_z=$ 0.5 state is higher than $q_z=$ 0.25. Thus ,the realised state with $q_z=$ 0.5 is supported by factors not captured by DFT calculations. Such factors could include subtle structural changes accompanying the first-order magnetic transition from Phase III to Phase I.
	
	In Phase I our neutron scattering experiments identified Fe-slabs with the Fe moments being parallel within the slabs stacked along $c$, while the moments of adjacent slabs are opposite. These slabs were highlighted already in the structural unit cell in Fig.\ref{Fig1} and their stacking in the magnetic unit cell is shown in Fig.~{\color{red}S6}. Such arrangement suggests strong (and predominantly) ferromagnetic intra-layer couplings combined with weaker antiferromagnetic inter-slab exchanges. We speculate that this competition at elevated temperatures leads to an incommensurate twist between the magnetic moments in the slabs.
	
	It is important to note that our SNP results unambiguously prove that the ICM order in Phase III is chiral, contrary to Ref. \cite{wu2018spin}, but in agreement with Ref. \cite{afshar2021spin}, while the population of two chiral domains is non-equal, the last being unexpected for a centrosymmetric structure. Similar nontrivial chiral properties were observed in YMn$_6$Sn$_6$~\cite{dally2021chiral} and we suspect that in both cases this population imbalance (and possibly tiny $\alpha$-component in \ce{Fe3Ga4}) is caused by strains or disorder discussed in Refs.\cite{mendez2015competing, wilfong2022altering}, which we observed as a reduction in crystal quality on rapid cooling at low temperatures. The tiny $\alpha$-component is an unexpected experimental outcome and its significance should still be understood. One possibility is that local variations in the structure allow for symmetry breaking between Fe-Fe bonds, which could generate a finite DMI.
	
	Our experiments show that the $q_z$ component changes abruptly with the field when passing from Phase III to Phase IV. This indicates that the change is likely related to the canting of the cycloidal TCS spiral, which is approximately linear in the field. The change of $q_z$ is significant (almost 20\%) within Phase IV for both reflections $\mathbf{Q}_1^{\mathrm{a}}$ and $\mathbf{Q}_2^{\mathrm{a}}$ (Fig.~{\color{red}S7e}). The increase is steeper at higher temperatures, suggesting the importance of thermal fluctuations. This behaviour is similar to YMn$_6$Sn$_6$~\cite{dally2021chiral}, wherein an in-plane magnetic field flops the helical spiral into the transverse conical one, that is a cycloid with a net moment component orthogonal to the plane of rotation. Concomitantly, a discontinuous change in $q_z$ is observed, albeit noticeably smaller than in \FG. For isotropic (Heisenberg) exchange systems the spiral pitch and subsequently the length of $\bf{q}$ is determined by frustration of exchange parameters and is independent of the spin-rotation plane. Another common parameter, single-ion anisotropy, also does not affect the length of $\bf{q}$. However, anisotropic exchange terms, such as $J^z S_1^zS_2^z$, can influence the magnitude of the propagation vector. In YMn$_6$Sn$_6$, where the Hamiltonian is much simpler, DFT calculations of $J_z$ quantitatively matches the observed jump of $q_z$. In \FG, the exchange network is too complicated to attempt such calculations. However, based on the changes in $q_z$ observed experimentally, we anticipate that anisotropic exchange interactions are present and likely are quite strong.
	
	As mentioned, when a magnetic field is applied to a spiral state within the spiral plane, the moments flop into the TCS state, which is a combination of a cycloid orthogonal to the field and a net moment component (see Fig.\ref{Fig1}a). The emergent magnetic field $\mathcal{B}$ within the TCS phase is zero and therefore no THE should exist. However, according to the fluctuation-based mechanism~\cite{ghimire2020competing}, the excitation probability of spin waves with opposite chirality is unequal, resulting in a non-zero $\mathcal{B}$ at finite $T$ within the TCS phase. This emergent magnetic field is linearly dependent on temperature and field, and quadratically dependent on net magnetisation. The experimental measured quantity is the topological Hall resistivity which is proportional to $\mathcal{B}$. As discussed above, \FG satisfies all the prerequisites for the topological Hall signal generated by the dynamical effects in Phase IV. Also, |$\rho_{\mathrm{THE}}$| is maximized in Phase V, indicating a possible non-trivial spin configuration. Since \FG belongs to a highly tuneable family of Fe-Ga alloys, our results open up attractive potential applications at room temperature. Some of these routes include, but are not limited to, emergent electromagnetic induction~\cite{yokouchi2020emergent}, uniaxial strain engineering, and dynamical Berry phase tuning~\cite{hirschberger2020topological} though precise chemical substitution, among others.
	
	\vspace{1cm}
	
	\noindent\textbf{\large METHODS}
	
	\noindent\textbf{Single crystal growth and characterisations}:\\ \FG~single crystal samples were grown by chemical vapor transport (CVT) method using iodine as the transport agent. Stoichiometric amount of Fe powder and Ga ingots, together with 250~mg of iodine were sealed inside a quartz ampoule of inner diameter 4~cm and length about 20~cm. The ampule was placed inside a two-zone furnace while the temperature of source and sink side were maintained at 550~$^\circ$C and 500~$^\circ$C, respectively. After two weeks of transport, the ampule was quenched to ice-cold water. The resultant crystals had a needle shape morphology with the long axis along $b$. All experiments mentioned in this article were performed on needle-type crystals, except for transport measurements which were carried out on a rectangular plate-like crystal.
	
	Quality and structure of all single crystalline specimens used in this study were verified using single crystal X-ray diffraction prior to other measurements. Single crystals were mounted on the goniometer head fitted with a cryo-loop. Data collection was performed on a Rigaku Synergy-I XtaLAB Xray diffractometer, equipped with a Mo micro-focusing source ($\lambda_{\mathrm{k\alpha}}$ = 0.71073 \AA) and a HyPix-3000 Hybrid Pixel Array detector (Bantam). Data reduction and absorption were carried out with CrysAlisPro and structure was solved and refined with ShelX package within OLEX2 software.\\
	
	\noindent\textbf{Magnetisation, susceptibility and electrical transport measurements}: \\ Magnetometry studies were performed using commercial 14T PPMS and MPMS3 from Quantum Design (QD). Except for measuring magnetisation under azimuthal rotation, needle-type single crystal was fixed on a quartz sample holder with the $b$-axis along the direction of magnetic field. Further for the $M(\psi)$ measurements, in order to access both $ac$ and $a^{\ast}b$  rotational planes, two types of rotors provided by QD were utilized. The same single crystal sample was fixed in the rotation center with a tiny amount of GE varnish.
	
	Electrical transport measurements were performed using the transport option of the same 14T PPMS. The plate-type single crystal was mounted on a sapphire plate with six contacts (two longitudinal and four transverse). For Hall effect measurements, magnetic field was stabilized at each point prior to data collection. \\
	
	\noindent\textbf{Single crystal neutron diffraction experiments}:\\ Several neutron diffraction experiments were performed on the single crystal constant-wavelength diffractometers ZEBRA, DMC (SINQ, PSI) and time-of-flight instrument WISH (ISIS, RAL). The dataset at 10 K in zero field was collected on ZEBRA in a cooling machine with the 4-circle setup. Diffraction experiments in magnetic field were performed at all three diffractometers using normal beam geometry. Two different 10~T vertical magnets were used at WISH and SINQ (DMC \& ZEBRA). The wavelengths used on DMC were 2.45 \AA\, and 4.52 \AA, while those on ZEBRA were 1.383 \AA\, and 2.31 \AA. The same needle-shaped single crystal (with [010] axis vertical) was used for all these experiments. The quality of the crystal was deteriorating on fast cooling but then recovering on heating to room temperature. This is the reason why some scans show multiple peaks originating from crystallites $\it circa$ 2 deg apart.\\
	
	\noindent\textbf{Small Angle Neutron Scattering (SANS) measurements}:\\ To track the temperature and field dependence of the ICM satellites, SANS studies were conducted using two instruments: D33 at the Institut Laue Langevin (ILL) and SANS-I at SINQ, PSI. For both experiments, a needle-type single crystal was aligned with the $b$-axis vertical (offset $\sim$ 0.6 deg) on an Al-plate, providing access to the ($h0l$) scattering plane.
	
	For the SANS experiments performed on D33, neutrons with wavelength $\lambda$ = 4.6 \AA~with a spread ($\Delta\lambda/\lambda$) of 10\% were used. The neutron beam was collimated at a distance of 2.8~m before the sample, while the main (side) detector was placed 2~m (1.2~m) behind the sample.
	
	For the second SANS experiment performed on SANS-I, neutrons with wavelength $\lambda$ = 5 \AA~and a similar wavelength spread were used. Neutron collimation distance and sample-detector distances were fixed at 4.5~m and 1.85~m, respectively. All SANS data analysis were performed using the GRASP software package~\cite{dewhurst2023graphical}. \\
	
	\noindent\textbf{Spherical neutron polarimetry (SNP) measurements}:\\ SNP experiment was carried out with the CRYOPAD setup on the hot neutron diffractometer D3 at the ILL. Similar to unpolarised neutron diffraction experiments, the needle-shaped single crystal was aligned with the  [010] axis vertical, giving access to the ($h0l$) scattering plane, see Fig.~\ref{Fig2}e and Fig.~{\color{red}S3}. Neutrons with the longest available wavelength of 0.843 \AA~selected by a Cu$_2$MnAl Heusler monochromator were used. In order to minimize beam depolarisation due to any external magnetic field, sample was positioned inside pair of cryogenically cooled Meissner shields. While the beam polarisation was controlled with a combination of nutator and precession coils, analysis of the outgoing neutron beam was performed by a field-polarized $^3$He spin filter cell, before finally being measured with a $^3$He detector. The final polarisation of the outgoing beam was corrected with respect to the cell efficiency. \\

	\bibliography{biblio}
	
	\vspace{10pt}
	
	\noindent\textbf{\large DATA AVAILABILITY}\\
	Data associated with this study are available from the corresponding authors upon reasonable request.\\
	
	\noindent\textbf{\large ACKNOWLEDGEMENTS}\\
	Swiss National Science Foundation (SNSF) projects, Sinergia Network \textquotedblleft NanoSkyrmionics\textquotedblright\, grant no. CRSII5\_171003 (P.R.B., V.U., A.M., $\&$ J.S.W.), 200021\_188707 (P.R.B., V.U., $\&$ J.S.W.), and 200020\_182536 (P.R.B. $\&$ O.Z.), are acknowledged for financial assistance. Additionally, P.R.B. acknowledges SNSF
	Postdoc.Mobility grant P500PT\_217697 for financial assistance. I.I.M. was partially supported by the National Science Foundation under Award No. DMR-2403804. We acknowledge beamtime allocation from PSI (20202357, 20230062, 20230040), ILL (DIR-241, 5-54-409), ISIS-RAL (2220587) and HLD-HZDR (member of the European Magnetic Field Laboratory (EMFL)) facilities. This work is based partly on experiments performed at the Swiss spallation neutron source SINQ, Paul Scherrer Institute, Villigen, Switzerland. We thank Dr. Wen Hua Bi for his assistance with the single crystal X-ray diffraction experiments. P.R.B. would like to thank Dr. Rinsuke Yamada, Prof. Naoya Kanazawa and Prof. Max Hirschberger for fruitful discussions. \\
	
	\noindent\textbf{\large AUTHOR CONTRIBUTIONS}\\
	P.R.B. and A.M. grew the single crystals and characterized them. P.R.B., I.\v{Z}., and Y.L. performed the magnetometry and transport measurements. High-field measurements were performed by P.R.B., Y.S., and O.Z. Neutron diffraction measurements were performed by P.R.B., V.U., F.O., P.M., L.K., and O.Z. Spherical neutron polarimetry experiments were performed by P.R.B., A.S., J.A.R., and O.Z. P.R.B., V.U., R.C. and J.S.W. performed the small angle neutron scattering measurements. Data analysis was performed by P.R.B. and O.Z. I.I.M. provided the theoretical interpretation. The manuscript was written by P.R.B. and O.Z., with input from I.I.M. All coauthors read and commented on the draft. P.R.B., V.U., and O.Z. conceived the project.\\
	
	\noindent\textbf{\large COMPETING INTERESTS}\\
	The authors declare no competing financial interests.

\end{document}